\documentclass[10pt]{article}
\usepackage{latexsym}
\usepackage{amsmath}
\usepackage{amssymb}
\title{Nonlinear Reformulation of Heisenberg's Dynamics}
\author{\begin{minipage}[t]{20ex}
  \begin{center}
   Martin Ziegler \\ \tt ziegler@upb.de
  \end{center}
\end{minipage} \and \begin{minipage}[t]{20ex}
  \begin{center}
  Benno Fuchssteiner \\ \tt benno@mupad.de
  \end{center}
\end{minipage}}
\date{University of Paderborn}
\newcommand{\IR}{\mathbb{R}}
\newcommand{\IC}{\mathbb{C}}
\newcommand{\IQ}{\mathbb{Q}}
\newcommand{\IP}{\mathbb{P}}
\newcommand{\IH}{\mathbb{H}}
\newcommand{\IF}{\mathbb{F}}
\newcommand{\IK}{\mathbb{K}}
\newcommand{\IG}{\mathbb{G}}
\newcommand{\IA}{\mathbb{A}}
\newcommand{\IB}{\mathbb{B}}
\newcommand{\IN}{\mathbb{N}}
\newcommand{\IS}{\mathbb{S}}
\newcommand{\IW}{\mathbb{W}}

\newcommand{\IV}{\mathbb{V}}
\newcommand{\IU}{\mathbb{U}}
\newcommand{\II}{\mathbb{I}}
\newcommand{\IO}{\mathbb{O}}
\newcommand{\Izero}{\IO}
\newcommand{\Ione}{\II}
\newcommand{\calA}{\mathcal{A}}
\newcommand{\calM}{\mathcal{M}}
\newcommand{\calN}{\mathcal{N}}
\newcommand{\calT}{\mathcal{T}}
\newcommand{\calS}{\mathcal{S}}
\newcommand{\calF}{\mathcal{F}}
\newcommand{\calB}{\mathcal{B}}
\newcommand{\calH}{\mathcal{H}}
\newcommand{\calU}{\mathcal{U}}
\newcommand{\calV}{\Gamma}
\newcommand{\calL}{\pmb{L}}
\newcommand{\Nabla}{\pmb{d}}

\newcommand{\Ideal}{\mathcal{J}}
\newcommand{\relation}{\setbox0=\hbox{$\equiv$}
  \mathrel{\hbox to 0pt{\hbox to \wd0{\raisebox{0.05ex}{$\scriptstyle\hbar$}\hss}\hss} \hbox to \wd0{\hss $\equiv$ \hss}}\!}
\newcommand{\defi}[1]{{\rm\textsf{#1}}}
\newcommand{\commut}[1]{\mycommut-{\left[ #1 \right]}}
\newtheorem{theorem}{Theorem}[section]
\newtheorem{metatheorem}{Meta-Theorem}[section]
\newtheorem{lemma}[theorem]{Lemma}
\newtheorem{definition}[theorem]{Definition}
\newtheorem{proposition}[theorem]{Proposition}
\newcommand{\COMMENTED}[1]{}
\DeclareRobustCommand{\qed}{%
\ifmmode 
\else \leavevmode\unskip\penalty9999 \hbox{}\nobreak\hfill
\fi
\quad\hbox{\leavevmode%
\hbox to.77778em{%
\hfil\vrule
\vbox to.675em{\hrule width.6em\vfil\hrule}%
\vrule\hfil}}}
\makeatletter
\newenvironment{iproof}{\par
\normalfont
\topsep6\p@\@plus6\p@ \trivlist
\item[\hskip\labelsep\itshape
\bf Proof\@addpunct{:}]\ignorespaces
}{%
\qed\endtrivlist
}\usepackage{a4}\newenvironment{proof}[1]{\par
\normalfont
\topsep6\p@\@plus6\p@ \trivlist
\item[\hskip\labelsep\itshape
\bf Proof of #1\@addpunct{:}]\ignorespaces
}{%
\qed\endtrivlist
}
\makeatletter
\def\overlie#1{\vbox{\m@th\ialign{##\crcr
      \hskip1.5pt \vrule depth0.5pt width4pt \hfill \vrule depth0.5pt
      width4pt \hskip1.5pt \crcr\noalign{\kern0pt\nointerlineskip}
      $\hfil\displaystyle{#1}\hfil$\crcr}}}
\def\underlie#1{\mathop{\vtop{\m@th\ialign{##\crcr
      $\hfil\displaystyle{#1}\hfil$\crcr\noalign{\kern0pt\nointerlineskip}
      \hskip1.5pt \vrule depth0.5pt width4pt \hfill \vrule
       depth0.5pt width4pt \hskip1.5pt
     \crcr\noalign{\kern0\p@}}}}\limits}
\newlength{\hoehe}
\def\mycommut#1#2{\setbox0=\hbox{#1}\setlength{\hoehe}{\ht0}
    \addtolength{\hoehe}{-2.0pt}\>
    \smash{\vtop{\ialign{##\crcr
           $\displaystyle{#2}$\crcr
           \noalign{\kern\hoehe\nointerlineskip}%
           \hfil\smash{$\scriptscriptstyle#1$}\crcr}}}
    \vphantom{\displaystyle{#2}}\>}
\makeatother
\newcommand{\LieBra}[1]{ \mathchoice
{\smash{\overlie{\underlie{\displaystyle\left|\!\left|\,#1\,\right|\!\right|}}}
 \vphantom{\displaystyle\left|\!\left|#1\right|\!\right|}}
{\smash{\overlie{\underlie{\textstyle\left|\!\left|\,#1\,\right|\!\right|}}}
 \vphantom{\textstyle\left|\!\left|#1\right|\!\right|}}
{\smash{\overlie{\underlie{\scriptstyle\left|\!\left|\,#1\,\right|\!\right|}}}
 \vphantom{\scriptstyle\left|\!\left|#1\right|\!\right|}}
{\smash{\overlie{\underlie{\scriptscriptstyle\left|\!\left|\,#1\,\right|\!
 \right|}}} \vphantom{\scriptscriptstyle\left|\!\left|#1\right|\!\right|}} }
\begin{document}
\maketitle

\begin{abstract}
A structural similarity between Classical Mechanics (CM) and
Quantum Mechanics (QM) was revealed by P.A.M.
Dirac in terms of Lie Algebras: while in CM the dynamics is
determined by the Lie
algebra of Poisson brackets on the manifold of scalar fields for
classical position/momentum observables $Q$/$P$, QM evolves (in
Heisenberg's picture) according to the formally similar Lie
algebra of commutator brackets of the corresponding operators:
$$\textstyle\frac{d}{dt}Q=\{Q,H\} \quad\frac{d}{dt}P=\{P,H\}
   \qquad\text{vs.}\qquad
  \frac{d}{dt}\IQ=\frac{i}{\hbar}\commut{\IQ,\IH} \quad
  \frac{d}{dt}\IP=\frac{i}{\hbar}\commut{\IP,\IH}$$
where $\IQ\IP-\IP\IQ=i\hbar$.
A further common framework for comparing CM and QM is the
category of symplectic manifolds.
Other than previous authors, this
paper considers phase space of Heisenberg's picture, i.e.,
the manifold of pairs of
operator observables $(\IQ,\IP)$ satisfying commutation relation.
On a sufficiently high algebraic level of abstraction ---
which we believe to be of interest on its own ---
it turns out that this approach leads to a
truly \emph{non}linear yet Hamiltonian reformulation of QM evolution.
\end{abstract}


\section{Introduction}
QM nowadays  is generally accepted as appropriate for describing very
small particles and their physical interactions and was put
into axiomatic form by von Neumann \cite{vNeumann}.
Since that time, much research has been spent on structural
similarities and differences between QM and Classical Mechanics
(CM). The usual approach is to propose a common mathematical
category where both theories fit into
and to then compare the additional axioms
satisfied by either theory. A simple example
are the commutation relations among Cartesian
position/momentum observables:
\begin{equation} \label{eqVertrel}
Q P-P Q=0 \quad\text{ in CM, } \qquad\qquad
  \commut{\IQ,\IP}:=\IQ \IP-\IP \IQ\overset{!}{=}i\hbar \quad\text{ in QM. ~}
\end{equation}
In the category of algebras, classical observables thus form a
\emph{commutative} one whereas  the quantum case becomes
commutative only as $\hbar$ tends to zero: one
consequence of Bohr's famous {Correspondence
Principle}.

Another category for comparing CM with QM arises from
the respective dynamics 
~ $\frac{d}{dt}A=\{A,H\}$~
and ~$\frac{d}{dt} \IA= \frac{i}{\hbar}\commut{\IA,\IH}$, 
~where we adopted the \emph{Heisenberg picture} that
observables $A(t)$ and $\IA(t)$ --- rather than states --- evolve
with time. Here $\{\,\cdot\,,\,\cdot\,\}$ denotes the
\emph{Poisson bracket} and $\commut{\,\cdot\,,\,\cdot\,}$ the
Commutator bracket. Since both brackets satisfy Jacobi's identity,
one arrives at a second common category for QM and CM: \emph{Lie
Algebras}, cf. \cite{DPG,Landsman}.

The present work adds to these perspectives a further
category by  proposing to consider \emph{Hamiltonian systems} in
the classical sense, a notion well-known in CM \cite{Mechanics},
however on abstract operator manifolds. Hamiltonian
systems have successfully been generalized from finite to infinite
dimensional manifolds \cite{Chernoff,Manifolds} and proven to be
the key to integrability of many difficult nonlinear partial
differential equations. \cite{KdV,Benno}.
We recall that if the generator $K$ of an evolutionary equation
\begin{equation} \label{eqBasic}
   \tfrac{d}{dt} u(t) \quad = \quad K\big(u(t)\big)
\end{equation}
is \emph{Hamiltonian} as a vector field\footnote{not to be
confused with the Hamiltonian \emph{operator}, i.e., the QM
observable for energy\ldots}, then conserved
quantities relate to symmetries in the sense of Noether.
Thus, in our approach, this relation holds also in
case of QM, thereby adding rich structural properties to QM.
Our work proceeds in three steps:
\begin{itemize}
\itemsep0pt
\item
  We turn the phase space of Heisenberg's picture into an
  (infinite dimensional) Banach manifold.
\item
  We consider on this manifold the generator of
  the Heisenberg dynamics and regard it as purely algebraic an object.
\item
  We show that this abstract object is a \emph{nonlinear} 
  Hamiltonian vector field in a sense similar to symplectic mechanics.
\end{itemize}
{\bf Related Work} considered,
in order to fit QM into the framework of Hamiltonian systems,
the 
Schr\"{o}dinger picture, i.e., an evolution on either the set of
vectors \cite{Chernoff,Kupershmidt} or on the set of rays (pure
states) in a Hilbert Space \cite{Heslot,Cirelli}. This
however, leads necessarily to \emph{linear} dynamics. Heisenberg's
dynamics, restricted to \emph{spin} space, was already
employed in \cite{Spin} to embed the
evolution of spin chains into a Hamiltonian framework.
The dynamic focused on in the
present work is again Heisenberg's but restricted to \emph{phase}
space, i.e., the equation
\begin{equation} \label{eqHeisenberg}
\frac{d}{dt}\,\bigg(\!\begin{array}{c}\IQ\\[0.5ex]\IP\end{array}\!\bigg)
\quad=\quad
\bigg(\begin{array}{c} \tfrac{i}{\hbar} \commut{\IQ,\IH} \\[0.7ex]
\tfrac{i}{\hbar} \commut{\IP,\IH} \end{array} \bigg)
\qquad =: \quad \IK(\IQ,\IP)
\end{equation}
which is assumed to take place on the set $\calM$ of all
tuples $(\IQ,\IP)$ of selfadjoint Hilbert space operators
satisfying,
in the sense of Weyl, 
the canonical commutation relation (\ref{eqVertrel}). The
question then is: In what sense can this in general nonlinear
dynamics be considered as a classical Hamiltonian flow?
In fact notice that for instance \emph{an}harmonic potential ~
$\IH=\IP^2 + \IQ^4$ ~ leads to a \emph{non}linear generator
$\IK(\IQ,\IP)=(2\IP,-4\IQ^3)$. This does not come to surprise as
classical Hamilton equations
\begin{equation} \label{eqHamilton}
 \frac{d}{dt}\,\bigg(\!\begin{array}{c}Q\\[0.5ex]P\end{array}\!\bigg)
\quad=\quad
\bigg(\!\begin{array}{c} \{Q,H\} \\[0.7ex] \{P,H\} \end{array} \!\bigg)
\qquad =: \quad K(Q,P)
\end{equation}
become nonlinear, too, for $H(Q,P)=P^2+Q^4$. 
Of course \cite{Mechanics}, the classical
Equation~(\ref{eqHamilton}) is always Hamiltonian, and our distant goal
is to show that also (\ref{eqHeisenberg}) is a classical Hamiltonian
flow, however on a new manifold of high dimension. Once that aim is
reached, then indeed we have revealed a genuine nonlinear
aspect of Quantum mechanics.
\\[0.5ex] {\bf Overview:}
As a first step, Section~\ref{secUnitary} turns the phase space
$\calM$ into an infinite-dimensional Banach manifold $\calM$.
However as $\IQ,\IP$ are not both bounded, the usual subtleties
arise \cite{Chernoff}:
weak vs. strong symplectic forms, issues of domains, and so on.
We circumvent these difficulties by considering
in Section~\ref{secAlgebra}, for
Hamiltonian operators that depend polynomially on phase space
variables (such as for example the above $\IH=\IP^2+\IQ^4$),
the induced generator of Heisenberg's dynamics (\ref{eqHeisenberg})
as purely \emph{algebraic} an object. 
Theorem~\ref{thAlgebra}, relying on
on a result in combinatorial algebra \cite{Ma},
formally justifies this identification.
In fact, 
\cite{Benno} revealed the
basic properties of usual Hamiltonian systems that relate
symmetries to conserved quantities and asserted the complete
integrability of so many important nonlinear flows
\cite{KdV,Gardner,Computer} to be expressible in 
\emph{algebraic} terms only, too.
Our major result 
(Section~\ref{secImplectic}) proves each such abstract
generator of Heisenberg's dynamics
to be Hamiltonian in this generalized algebraic sense.
But let us start with a brief review of some basic notions from 
dynamical systems and differential geometry.


\section{Manifolds, Flows, 
and Integrability}
\label{secManifold}
The present section contains a short introduction to
infinite dimensional manifolds, differential equations
thereon, and the impact of Hamiltonian
generators to the integrability of such equations.
This presentation closely follows \cite{Manifolds,Benno}.

\begin{definition} \label{defDerivative}
Let $E,F$ denote Banach spaces
and $U$ some open subset of $E$.
A function $f:U\to F$ is called \defi{differentiable}
at $x\in U$ if there exists a continuous linear map
$T=T[\cdot]:E\to F$ such\footnote{%
We adopt the convention that arguments
entering linearly are written in square brackets.%
} that
$$   \| f(x+v) - f(x) - T[v] \| / \|v\| \;\to\; 0
\qquad \text{ as } \quad  E\ni v\to 0 \enspace . $$
In that case, $T$ is unique and denoted $T=f'(x)$.
\end{definition}

More generally, it usually suffices for $E,F$ to be
\emph{locally convex Hausdorff} rather than Banach vector spaces
and $f$ to be \emph{Hadamard}-differentiable, cf. \cite{Benno}.
This allows for a variety of manifolds, e.g., modeled
over the space of rapidly decreasing functions $\calS$ or equipped
with some inductive limit topology. In fact for the following
considerations, the exact notion of differentiability is of minor
importance as long as it satisfies properties expressible
in purely algebraic terms:
\begin{itemize}
\itemsep0pt
\item
  chain rule for differentiation of composite functions;
\item
  product rule for differentiation;
\item
  symmetry of second derivatives;
\item
  and (occasionally) the implicit function theorem.
\end{itemize}
By means of charts, differentiability is then
carried over to functions $f:\calM\to\calN$ on manifolds $\calM$
and $\calN$ modeled over $E$ and $F$, respectively; see
\cite{Manifolds}. In particular, for a differentiable scalar field
$H:\calM\to\IR$ and $u\in\calM$, $I'(u)$ is a continuous linear
mapping from tangential space $\calT_u\calM$ to $\IR$, i.e., a
covector $\Nabla I(u)=I'(u)\in\calT^*_u\calM$, and $\Nabla
I:\calM\to\calT^*\calM$ a covector field.

For a vector field $K:\calM\to\calT\calM$, $\calM\ni u\mapsto
K(u)\in\calT_u\calM$, consider the
general-type evolutionary equation (\ref{eqBasic}). Its
solution $t\mapsto u(t)$ for given initial value $u(0)$ is called
a \emph{flow} or \emph{integral curve}. 
In the sequel, in order not to obscure the main ideas  by
technical details, we
shall, for simplicity, assume that $K$ is such
that this solution always exists, is unique, and well-behaved
(e.g., $C^k$). Similarly, fields are assumed to be smooth
enough such that all occurring derivatives make sense.

Indeed, the basic properties that lead to infinitely many
conserved quantities and symmetries for (\ref{eqBasic}) are
usually stated in purely algebraic terms
\cite{KdV,Benno}\footnote{See also Section~\ref{secAlgebra} of the
present work.} 
and a sound analytical foundation is given only later by
supplying $\calM$ with a suitable topology.
Observe that equations of the form (\ref{eqBasic}) cover
evolving physical systems ranging from simple pendulum
$$ \tfrac{d^2}{dt^2} \varphi + \sin\varphi=0, \qquad\text{ i.e. }\quad
  \calM:=\IR^2, \quad u:=(\varphi,\dot\varphi), \quad
  K(u):=(\dot\varphi,-\sin\varphi) $$
up to complicated partial differential equations
like the one due to Korteweg and de Vries describing
one-dimensional long water waves $u=u(x,t)$
\begin{equation} \label{eqKdV}
  \partial_t u \quad = \quad
    6 u \cdot \partial_x u + \partial^3_x u
    \quad =: \quad K\big(u)
\end{equation}
on some suitable manifold of
functions in one real variable.
Solving the latter used to be inherently difficult,
even numerically, due to its nonlinearity. The celebrated
breakthrough in \cite{KdV} was to show that (\ref{eqKdV})
possesses an infinite number of conserved quantities
related to symmetries by virtue of (a variant of) Noether's theorem.

\begin{definition} \label{defSymmetry}
A \defi{conserved quantity} for (\ref{eqBasic}) is a
scalar field $I:\calM\to\IR$ such that
$$ \Nabla I[K] : \calM\to\IR, \qquad u\mapsto I'(u)[K(u)] $$
is identically zero.
A \defi{symmetry} is a vector field $G:\calM\to\calT\calM$ such that
the following function vanishes identically:
\begin{equation} \label{eqSymmetry}
  \LieBra{K,G}:\calM\to\calT\calM, \qquad
    \LieBra{K,G}(u) := G'(u)[K(u)]-K'(u)[G(u)]
\end{equation}
\end{definition}

Notice that $I$ is a conserved quantity iff, for each flow
$t\mapsto u(t)$ of (\ref{eqBasic}), $t\mapsto I\big(u(t)\big)$
remains constant; cf. \textsc{Proposition~3.4.2} in \cite{Mechanics}.
Similarly, $G$ is a \emph{symmetry} iff
the one-parameter groups of flows induced by $K$ and $G$, respectively,
commute; cf. e.g. \textsc{Observation~2.2} in \cite{Benno}
or \textsc{Theorem, p.150} in \cite{Manifolds}.
We remark that $\LieBra{\cdot,\cdot}$ turns the set of vector fields
into a Lie Algebra. Indeed, $G'[K]-K'[G]$ is chart independent
(which, e.g., $G'[K]$ only is not), antisymmetric, and satisfies
Jacobi's identity (due to chain rule of differentiation and
symmetry of second derivatives).

Conserved quantities permit to reduce the dimension of the manifold
under consideration.
cf. \textsc{Exercise 5.2H} in \cite{Mechanics} or
\textsc{p.125} in \cite{Chernoff}. This explains
the notion \defi{integrable} for systems
(\ref{eqBasic}) that exhibit a \emph{complete} collection of
conserved quantities/symmetries, see \textsc{Definition~5.2.20}
in \cite{Mechanics}. It was therefore quite celebrated
when researchers discovered the famous Korteweg-de Vries Equation
(\ref{eqKdV}) to be integrable, that is, soluble in a rather
explicit and practical sense \cite{KdV,Marsden,Computer}.
As we now know, integrability also applies to a vast
number of other important nonlinear partial differential
equations such as, e.g., Gardner's, Burger's, nonlinear
Schr\"{o}dinger, and sine Gordon.
Furthermore, abstract integrability turned out to be closely
related to Hamiltonian structure \cite{Gardner,Benno} in
a purely algebraic sense.
One important aspect of this relation is expressed by the
following well-known (variant of a) result due to Emmy Noether:

\begin{metatheorem} \label{thNoether}
Let $K$ denote a Hamiltonian vector field on $\calM$. Then, to
every conserved quantity of (\ref{eqBasic}), there corresponds a
symmetry.
\end{metatheorem}

Here, a vector field $K:\calM\to\calT\calM$ is called
\emph{Hamiltonian} if some \emph{symplectic} 2-form
$\omega:\calT\calM\times\calT\calM\to\IR$,
identified with $\omega:\calT\calM\to\calT^*\calM$,
maps it ($K$) to the
gradient of a scalar field\footnote{usually
the \emph{energy} functional associated with the system\ldots}
$H:\calM\to\IR$, that is, an exact covector field:
\begin{equation} \label{eqHamiltonian}
  \omega\circ K : \calM\to\calT^*\calM \qquad \overset{!}{=} \qquad
   \Nabla H: \calM\to\calT^*\calM \enspace ;
\end{equation}
compare., e.g., \textsc{p.12} in \cite{Chernoff}, \textsc{Definition~5.5.2}
in \cite{Mechanics}, or \textsc{Section~VII.A.2} in \cite{Manifolds}.
Also notice that one of the requirements for $\omega$
to be symplectic is that at each $u\in\calM$,
the linear map $\omega(u):\calT_u\calM\to\calT^*_u\calM$
has a continuous inverse. Therefore, $K$ is uniquely
determined by $H$ and $\omega$; more precisely,
$K=\omega^{-1}[\Nabla H]$ for linear antisymmetric
$\omega^{-1}(u):\calT^*_u\calM\to\calT_u\calM$.

As was later observed, the proof
of Noether's theorem in fact exploits only algebraic properties
(e.g., symmetry of second derivatives or Jacobi's identity).
Indeed for $K:\calM\to\calT\calM$, the mappings
in Definition~\ref{defSymmetry} ---
$\calL_K:I\mapsto \Nabla I[K]$ on the set $\calF$ of scalar fields and
$\calL_K:G\mapsto \LieBra{K,G}$ on the set $\calV$ of vector fields
 --- as well as their canonical extensions to the set $\calV^*$ of
covector fields and to tensor fields of higher rank
according to \textsc{Section~III.C.2} in \cite{Manifolds},
are \emph{derivations} in the algebraic sense;
cf. \textsc{Proposition, p.148} in \cite{Manifolds}.
In fact, \textsc{Section~4} of \cite{Benno} and
\textsc{Section~2} of \cite{Coupling} gradually stripped
down the prerequisites of Theorem~\ref{thNoether}
and found it to holds on a far more abstract level:

\begin{definition}   \label{defNew}
Let $\big(\calV,\LieBra{\cdot,\cdot}\big)$ denote a Lie algebra
and $\calF$ a vector space ---
called (abstract) \defi{vector} and \defi{scalar fields},
respectively. Suppose that, for each $K\in\calV$,
$\calL_K:\calF\to\calF$ is such that $K\mapsto\calL_K$ is
injective and a Lie algebra homomorphism, i.e., satisfies
\begin{equation} \label{eqHomomorphism}
 \calL_K \calL_G - \calL_G \calL_K \quad = \quad \calL_{\LieBra{K,G}}
  \enspace .
\end{equation}
For $H\in\calF$ call the linear map
$$ \Nabla H:\calV\to\calF,
  \qquad K\mapsto \calL_K(H) \quad=:\quad \Nabla H[K] $$
a (abstract) \defi{gradient} or \defi{covector field};
the set of all of them being denoted by $\calV^*$.

Now extend, similarly to
\textsc{Section~III.C.2} in \cite{Manifolds},
\defi{Lie derivative} $\calL_K$ from scalar to
vector and covector fields $G\in\calV$ and $\Nabla H\in\calV^*$,
respectively
\begin{equation} \label{eqCovector}
  \calL_K G:=\LieBra{K,G}, \qquad
   (\calL_K\Nabla H)[G]:=\calL_K(\Nabla H[G])-\Nabla H[\calL_K G]
\end{equation}
and finally to tensors of higher rank; cf. \textsc{Equation~(2.4)}
in \cite{Coupling}.
A linear antisymmetric mapping
$\Theta:\calV^*\to\calV$, is called \defi{Noetherian} if
$\calL_{\Theta[\Nabla H]}(\Theta)=0$ for all $\Nabla H\in\calV^*$.
Call $K\in\calV$ \defi{Hamiltonian} if
$K=\Theta[\Nabla H]$ for some $H\in\calF$.
\end{definition}

In the usual setting, $\calF$ is the commutative algebra of
(sufficiently well-behaved, at least $C^1$) scalar fields on
manifold $\calM$, $\calV$ the set of vector fields
on $\calM$, and $\calV^*$ the set of all (conventional)
gradients, i.e., a proper subset of all continuous
linear local functionals on $\calV$.
But Definition~\ref{defNew} allows for $\calL_K$ to
operate also \emph{non}locally on $\calF$;
in fact, no underlying manifold is required at all
as long as $\calF$, $\calV$, and $\calL_K$ satisfy
algebraic conditions similar to conventional scalar/vector
fields and Lie derivatives on some $\calM$.
Concerning the notion of a {Noetherian} operator:
\textsc{Theorem~4.5} in \cite{Benno} contains six equivalent
characterizations for antisymmetric linear $\theta:\calV^*\to\calV$
to satisfy this requirement.
They reveal that 
"{\em
 loosely speaking, $\theta$ has the algebraic behavior of the
 inverse of a \emph{symplectic} operator
 }" \cite[\textsc{p.223}]{Benno}.
In other words: rather than imposing (regularity and other)
conditions on a $(2,0)$-tensor $\omega$
such that $K=\omega^{-1}[\Nabla H]$,
the last part of Definition~\ref{defNew} considers
$K=\theta[\Nabla H]$ and imposes conditions
on the $(0,2)$-tensor $\theta$ \emph{directly}.
It thus generalizes Equation~(\ref{eqHamiltonian})
while avoiding explicit nondegeneracy requirements which,
in infinite dimension, become
subtly ambiguous (injective/surjective/bijective) anyway.
Relevance of these dramatic generalizations is
illustrated by, among others\footnote{e.g., \textsc{Observations~5.2}
and \textsc{5.3} in \cite{Benno}\ldots}, the following
\begin{theorem} \label{thNew}
Let $K=\Theta[\Nabla H]\in\calV$ be Hamiltonian and $I\in\calF$
s.t. $\Nabla I[K]=0$~ (i.e., an \defi{abstract conserved quantity}).
Then, $G:=\Theta[\Nabla I]\in\calV$ satisfies
$\LieBra{K,G}=0$~ (i.e., is an \defi{abstract symmetry}).
\end{theorem}
\begin{iproof}
See \textsc{Theorem~3.3} in \cite{Benno} 
or Appendix~\ref{subsecNew} in {\tt quant-ph/0210198}.
\end{iproof}

This gives late justification why the purely formal manipulations
in \cite{KdV} actually did yield infinitely many conserved
quantities in the conventional sense. More precisely,
Definition~\ref{defNew} and Theorem~\ref{thNew} permit
to separate algebraic conditions from analytic ones; the latter are,
for the nonlinear partial differential equations already mentioned,
usually taken care of later by choosing as manifold $\calM$ some
appropriate function space with a suitable topology.


\section{Phase Space Manifold of Heisenberg's Picture} \label{secUnitary}
Consider a QM system with one spacial degree of freedom
and let $\calH$ denote some infinite-dimensional separable Hilbert
space. In Heisenberg's picture, phase space $\calM$ consists of all
pairs $(\IQ,\IP)$ of selfadjoint operators on $\calH$
satisfying, in the sense of Weyl, the canonical commutation
relation (\ref{eqVertrel}).

The below considerations are easily generalized to
QM systems with $f>1$ spacial degrees of freedom,
where phase space consists of $2f$-tuples
$(\IQ_1,\IP_1$,\; $\IQ_2,\IP_2$,\; \ldots,\; $\IQ_f,\IP_f)$
of selfadjoint operators satisfying
\begin{equation} \label{eqVertrel2}
\IQ_k\IQ_l=\IQ_l\IQ_k, \qquad \IP_k\IP_l=\IP_l\IP_k, \qquad
\IQ_k\IP_l-\IP_l\IQ_k=i\hbar\delta_{kl} \enspace .
\end{equation}
It is merely for notational convenience that in this section we will
focus on the case $f=1$ and show how to turn the set $\calM$ into
a Banach manifold in the sense of
\cite[\textsc{Section~VII.A.1}]{Manifolds}.

To this end recall von Neumann's celebrated result that
each such pair $(\IQ,\IP)$ is
unitarily equivalent to a fixed pair $(\IQ_0,\IP_0)$;
cf. e.g. \textsc{Theorem~4.3.1} in \cite{Putnam}
or \textsc{Theorem~VIII.14} in \cite{Reed}:
\qquad $\IQ=\IU\IQ_0\IU^* \quad\wedge\quad \IP=\IU\IP_0\IU^*$
\qquad for some unitary  $\IU$.
We explicitly disallow multiplicities/direct sums
because systems with \emph{one} degree of freedom
correspond to \emph{irreducible} representations of
Schr\"{o}dinger couples.
Since conversely, every pair $(\IU\IQ_0\IU^*,\IU\IP_0\IU^*)$
does satisfy (\ref{eqVertrel}), it suffices to consider the
set $\calU(\calH)$ of all unitary operators on $\calH$.
We assert that this set actually indeed \emph{is} a manifold.

\begin{theorem}\footnote{%
Due to a page limit, proofs
had to be moved into an appendix available at \tt quant-ph/0210198.}
The set $\calU(\calH)$ of unitary operators on a
separable \label{thUnitary}
Hilbert space $\calH$ is a real $C^\infty$ Banach manifold.
\end{theorem}


\section{Polynomials in Operator-Variables} \label{secAlgebra}
In CM, the phase
space\footnotemark\addtocounter{footnote}{-1} manifold
$\calM$ consists of canonical position/momentum
variables $(Q,P)$, and the Hamilton function depends
smoothly (say, rationally or polynomially) on these variables.

In Heisenberg's picture of QM,
the phase space\footnote{of systems with one spacial degree of freedom}
manifold $\calM$ consists of canonical position/momentum
observables, that is, of tuples $(\IQ,\IP)$ of selfadjoint
Hilbert space operators satisfying,
in the sense of Weyl,
commutation relations (\ref{eqVertrel}).
This time, the Hamilton function
is an operator-valued function of such tuples
namely a Hamiltonian operator like $\IH(\IQ,\IP)=\IQ^4+\IP^2$.

We will in the sequel focus on Hamiltonians depending
\emph{polynomially} on $\IQ$ and $\IP$, and the aim
of this section is to make this notion mathematically sound.
In algebra, polynomials $p\in\IC[Q,P]$ in two variables are
of course well-defined objects. The following four
definitions are known to be equivalent:

\begin{definition} \label{defPoly1}
The set $\IC[X_1,\ldots,X_m]$ of \defi{polynomials}
over $\IC$ in $m$ (commutating) variables is
\begin{enumerate}
\itemsep0pt
\item
  the free commutative $\IC$-algebra generated by $\{X_1,\ldots,X_m\}$;
\item
  the set of finite unbounded sequences (namely the coefficients
  preceding monomials) with convolution as product;
\item
  the smallest family of mappings $\hat p:\IC^m\to\IC$ containing
  constants and projections
  $(x_1,\ldots,x_m)\mapsto x_j$ and being closed under addition
  and multiplication;
\item
  the collection of all differentiable mappings $\hat p:\IC^m\to\IC$
  for which differentiation is
  nilpotent, i.e. for some $k$, the $k$-th derivative vanishes\footnote{%
  Recall that, according to complex analysis, any differentiable
  function on $\IC$ is necessarily $C^\infty$.}.
\end{enumerate}
\end{definition}
While appropriate for classical (i.e., commuting) observables,
this type of polynomials however does not
reflect the non-commutativity in general exhibited by
quantum observables.
Instead consider a definition of polynomials in
\emph{non}commuting variables similar to a):

\begin{definition} \label{defPoly2}
The set $\IC\langle X_1,\ldots,X_m\rangle$ of
\defi{polynomials} over $\IC$ in non-commuting variables
is the free non-commutative (but
associative and distributive) $\IC$-algebra generated
by $\{X_1,\ldots,X_m\}$.
A \defi{monomial} in $\IC\langle X_1,\ldots,X_m\rangle$
of \defi{degree} $d$ is of the form $\prod_{n=1}^d X_{k_n}$
with $k\in \{1,\ldots,m\}^d$.
\end{definition}

As each such polynomial is obviously a linear combination of
finitely many monomials and vice versa, one easily obtains
an equivalent characterization in terms of coefficient sequences
similar to Definition~\ref{defPoly1}b); the convolution just
doesn't look as nice any more. But how about analogues
to c) and d), that is, a way to identify polynomials with
certain differentiable mappings on \emph{quantum} observables?

Of course for some $\IC$-algebra $\calA$ --- such as the
set of selfadjoint linear operators on some Hilbert space ---
every $p\in\IC\langle X_1,\ldots,X_m\rangle$ gives rise to a
mapping $\hat p:\calA^m\to\calA$ where
$\hat p(A_1,\ldots,A_m)$ is defined by substituting
$X_j$ with $B_j$. But is $\hat p$ differentiable?
Moreover, $p\mapsto\hat p$ is a homomorphism;
but in order to \emph{identify} $p$ with $\hat p$, this
homomorphism should be injective! As one can easily imagine,
this heavily depends on $\calA$ to be sufficiently rich;
for example the two polynomials in two non-commuting variables
$p,q\in\IC\langle X,Y\rangle$ with $p:=X\cdot Y$ and $q:=Y\cdot X$
differ whereas $\hat p$ and $\hat q$ agree on $\calA:=\IC$.
Similarly, $p:=(X\cdot Y-Y\cdot X)^2\cdot Z$ and
$q:=Z\cdot(X\cdot Y-Y\cdot X)^2$ from $\calA:=\IC\langle X,Y,Z\rangle$
satisfy, according to \emph{Hall's Identity},
$\hat p=\hat q$ on the algebra $\calA$ of $2\times 2$ matrices.

This section's main result asserts that already
the set of compact symmetric linear operators on
infinite-dimensional Hilbert space $\calH$
is sufficiently rich to identify polynomials
with polynomial mappings. Furthermore on bounded
linear operators, these polynomial mappings are
differentiable.

\begin{theorem} \label{thAlgebra}
Let $\calH$ denote some separable infinite-dimensional Hilbert space,
\linebreak[4] $p,q\in\IC\langle X_1,\ldots,X_m\rangle$.
\begin{enumerate}
\itemsep0pt
\item
  Let $\calA$ be a set of linear operators on $\calH$
  containing at least the compact symmetric ones.
  Then $\hat p\big|_{\calA^m}=\hat q\big|_{\calA^m}$ implies $p=q$.
\item
  Let $\calB$ be the Banach algebra of bounded linear operators
  on $\calH$. Then $\hat p:\calB^m\to\calB$ is differentiable.
\item
  Its derivative
  ${\hat p}'(\IA_1,\ldots,\IA_m)[\IV_1,\ldots,\IV_m]$
  is the $\widehat{~}$-transform of some unique polynomial
  in $2m$ non-commuting variables and thus differentiable as well.
\end{enumerate}
\end{theorem}

We may thus --- and will from now on --- use
\emph{polynomial} (in non-commuting variables) and
\emph{polynomial mapping} synonymously.
By virtue of part c), every polynomial is $C^\infty$,
and one may write $p'(X_1,\ldots,X_m)[V_1,\ldots,V_m]\in
\IC\langle X_1,\ldots,X_m;V_1,\ldots$, $V_m\rangle$
for the derivative of $p\in\IC\langle X_1,\ldots,X_m\rangle$.

As next step one has to take into account the commutation relation
(\ref{eqVertrel}) satisfies by quantum phase space observables.
Indeed, the polynomials $p:=QP-PQ$ and $q:=i\hbar$ are different
in $\IC\langle P,Q\rangle$
whereas for position/momentum observables $\IQ/\IP$ it holds
$\IQ\IP-\IP\IQ=i\hbar$. One therefore wants to identify
polynomials $\sum_{k=1}^K p_k\cdot(QP-PQ-i\hbar)\cdot q_k$
in $\IC\langle Q,P\rangle$ with $0$
while maintaining the structure of an algebra
such as being closed under addition and multiplication.

Fortunately, exactly this is offered by the
\emph{quotient algebra}
\begin{equation} \label{eqQuotient}
\begin{gathered}
\IC\langle Q,P\rangle/\Ideal \;:=\;
   \Big\{ p/\Ideal : p\in\IC\langle Q,P\rangle \Big\},
   \qquad  p/\Ideal:=\{p+q : q\in\Ideal \}, \\
      (p_1/\Ideal) + (p_2/\Ideal) \cdot (p_3/\Ideal) \;:=\;
      (p_1+p_2\cdot p_3)/\Ideal
\end{gathered}
\end{equation}
where $\Ideal$ denotes some appropriate ideal.
Recall that a (two-sided) ideal is a subset of an algebra which
is closed under addition and closed under
multiplication (both left and right) with arbitrary elements
not only from $\Ideal$ but from the whole algebra.
In our case, take the ideal spanned by
$QP-PQ-i\hbar\in\IC\langle Q,P\rangle$,
that is, the smallest\footnote{reflecting that \emph{no other}
identifications than $QP-PQ=i\hbar$ are to be made\ldots}
ideal containing $QP-PQ-i\hbar$; explicitly:
$$ \textstyle \Ideal \quad=\quad
\Big\{ \sum\nolimits_{k=1}^K \; p_k\cdot (QP-PQ-i\hbar) \cdot q_k \;:\;
K\in\IN_0, \;p_k,q_k\in\IC\langle Q,P\rangle
\Big\} \enspace . $$
Considering $\IC\langle Q,P\rangle/\Ideal$ rather than
$\IC\langle Q,P\rangle$, it now holds
$\IQ\IP-\IP\IQ=i\hbar$ for elements $\IQ:=Q/\Ideal$ and $\IP:=P/\Ideal$.
Let us abbreviate $\IC\langle\IQ,\IP\rangle:=\IC\langle Q,P\rangle/\Ideal$
and remark that $\IQ,\IP$ like $Q,P\in\IC\langle Q,P\rangle$,
are in some sense not `variables' but very specific
and purely algebraic objects.
On the other hand recall that by virtue of the above considerations,
each element $\IH$ from $\IC\langle Q,P\rangle/\Ideal$ does give
rise to and can be identified with a mapping $\hat\IH$ on the
phase space manifold of all pairs of quantum position/momentum
observables. We are thus led to the following

\begin{definition} \label{defScalar}
The algebra $\calF:=\IC\langle\IQ,\IP\rangle$ is called
the set of \defi{abstract scalar fields} on quantum
phase space.
\end{definition}


\section{Hamiltonian Heisenberg's Dynamics} \label{secImplectic}
We will now use and extend the purely algebraic approach
from Section~\ref{secAlgebra} to prove 
that the generator of Heisenberg's dynamics is Hamiltonian
in the 
sense of Definition~\ref{defNew}.

A first step, the set $\calF$ of abstract scalar fields
has already been introduced in Definition~\ref{defScalar}.
This formalized the class Hamiltonian operators of interest:
polynomials $\IH=\IH(\IQ,\IP)$ in phase space variables
$(\IQ,\IP)=u$.
According to Definition~\ref{defNew}, next we need is
a Lie algebra $\calV$ to serve as (abstract) vector
fields, i.e., containing generators $\IK=\IK(u)$ of
a dynamics (\ref{eqBasic}) on phase space manifold
$\calM=\big\{ (\IQ,\IP) \;:\; \IQ\IP-\IP\IQ=i\hbar\big\}$.
In order for corresponding flows $t\mapsto u(t)=\big(\IQ(t),\IP(t)\big)$
to stay on $\calM$, $\IK=(\IK_Q,\IK_P)$ must not alter commutation relation
(\ref{eqVertrel}), i.e.,
\vspace*{-1ex}
\begin{eqnarray} \label{eqVector}
0 &\overset{!}{=}& \tfrac{d}{dt} 0
  \quad=\quad \tfrac{d}{dt} \big(\IQ(t)\IP(t)-\IP(t)\IQ(t)-i\hbar\big)
\nonumber \\ &\overset{(*)}{=}&
  \big(\tfrac{d}{dt} \IQ(t)\big)\IP(t)+\IQ(t) \big(\tfrac{d}{dt}
  \IP(t)\big) - \big(\tfrac{d}{dt}\IP(t)\big)\IQ(t)
   - \IP(t)\big(\tfrac{d}{dt}\IQ(t)\big)
\nonumber \\ &\overset{(\ref{eqBasic})}{=}&
  \IK_Q\IP+\IQ\IK_P - \IK_P\IQ-\IP\IK_Q
\quad=\quad \commut{\IK_Q,\IP}-\commut{\IK_P,\IQ}
\vspace*{-1ex}
\end{eqnarray}
where at $(*)$ we used the product rule of differentiation
and exploited that $\hbar$ does not vary over time.
The algebraic excerpt of these considerations is subsumed in

\begin{definition}  \label{defVector}
The set of \defi{abstract vector fields} on quantum phase space is
$$ 
 \calV \quad=\quad \big\{
  \IK=(\IK_Q,\IK_P) \;:\;
   \IK_Q,\IK_P\in\IC\langle\IQ,\IP\rangle \;,
    \; \commut{\IK_Q,\IP}=\commut{\IK_P,\IQ} \big\}
$$ 
For ~$\IK=(K_Q/\Ideal,K_P/\Ideal)\in\calV$~ and ~$\IH=H/\Ideal\in\calF$~
with ~$K_Q,K_P,H\in\IC\langle Q,P\rangle$,
$$  \calL_{\IK} \IH \quad := \quad \big(H'[(K_Q,K_P)]\big)/\Ideal
  \quad\in\quad \calF $$
denotes the \defi{Lie derivative} of $\IH$ with respect to $\IK$.
Finally equip $\calV$ with the bracket
$$ \LieBra{\IK,\IG} \quad:=\quad
  \big( \calL_{\IK} \IG_Q-\calL_{\IG}\IK_Q ,
        \calL_{\IK} \IG_P-\calL_{\IG}\IK_P \big) \enspace . $$
\end{definition}

These structures indeed comply with
the requirements in Definition~\ref{defNew}:

\begin{theorem} \label{thWelldefined}
$\calL_{\IK}\IH$ in Definition~\ref{defVector} is well-defined.
$\LieBra{\cdot,\cdot}$ 
 turns the vector fields $\calV$ into a Lie algebra.
$\IK\mapsto\calL_{\IK}$ is an injective Lie algebra homomorphism.
\end{theorem}

Remember our goal to find a Hamiltonian formulation for
the generator
 $\IK:=\big( \tfrac{i}{\hbar}\commut{\IQ,\IH} ,
  \tfrac{i}{\hbar}\commut{\IP,\IH} \big)$
of Heisenberg's dynamics on phase space
(\ref{eqHeisenberg}).
To this end, one needs
some Noetherian $\Theta:\calV^*\to\calV$
such that $\IK$ as $\Theta[\Nabla\IH]$.
Compare this to the situation in CM (\ref{eqHamilton})
where the generator is well-known \cite{Mechanics}
to have a Hamiltonian formulation $K=\theta[\Nabla H]$
by means of
$\theta:=
\Big(\!\!\begin{array}{r@{\;\:}r}
 0 &+1\\ -1 & 0 \end{array}\!\!\Big)$.
Indeed when identifying gradients (i.e., covectors)
with vectors, $\theta$ is obviously linear, antisymmetric, and
even constant hence $\calL_K \theta=0$ for any $K$. Moreover the well-known
\emph{Darboux theorem} states that conversely \emph{every}
symplectic tensor on a $2f$-dimensional manifold is of the form ~
$\Big(\!\!\begin{array}{r@{\;\:}r}
 \Izero_f &+\Ione_f\\ -\Ione_f & \Izero_f \end{array}\!\!\Big)$
~ at least locally, where $\Izero_f$, $\Ione_f$ denote
the zero and identity $(f\times f)$-matrix, respectively.

Of course on \emph{in}finite-dimensional manifolds, covectors from
$\calV^*$ cannot in general be identified with vectors from $\calV$.
But still the following consideration conveys the idea that turns
out to carry over to our algebraic setting where $\Theta:\calV^*\to\calV$
need not be bijective. To this end observe that a two-dimensional
covector $w^*$ on $\IR^2$, i.e., a linear function $w^*:\IR^2\to\IR$,
is identified with a vector $w\in\IR^2$ via
$$  w \quad=\quad
   \big( \; w^*[(1,0)] \;,\; w^* [(0,1)]
     \; \big) \quad \in \quad \IR^2 \enspace , \text{i.e.,} $$
by evaluating $w^*$ at
arguments $(1,0)$ and $(0,1)$ forming the canonical basis for $\IR^2$.

\begin{definition} \label{defSymplectic}
For abstract covector field $\IW^*\in\calV^*$, let
$$ \Theta[\IW^*] \quad := \quad
   \bigg(\!\begin{array}{rr} 0 & +1 \\[0.7ex] -1 & 0
  \end{array}\!\bigg) \cdot
     \bigg( \begin{array}{c} \IW^*[(1/\Ideal,0/\Ideal)] \\[0.7ex]
       \IW^*[(0/\Ideal,1/\Ideal)] \end{array}
     \bigg) \quad \in \quad \calV $$
\end{definition}

\begin{theorem}  \label{thImplectic}
$\Theta:\calV^*\to\calV$ is well-defined and Noetherian.
Furthermore for each abstract scalar field $\IH\in\calF$,
the (thus Hamiltonian) vector field $\Theta[\Nabla\IH]$
coincides with $\IK$ according to (\ref{eqHeisenberg}).
\end{theorem}

Although $\Theta$ resembles the classical $\theta$, the proof in
Appendix~\ref{subsecImplectic} proceeds entirely different.
In fact already the antisymmetry of $\Theta$ is far from obvious
and heavily relies on $\calF$ being the quotient algebra
with respect to $\Ideal$.

\section{Conclusion} \label{secConclusion}
We showed that, for Hamiltonian operators that
depend polynomially on observables $\IQ$ and $\IP$,
Heisenberg's dynamics on phase space is Hamiltonian
at least in an \emph{abstract algebraic} sense.
This constitutes an important step and in fact can serve 
as a guide towards a Hamiltonian formulation of QM dynamics
as \emph{analytical} flow on a \emph{concrete} manifold like
the one considered in Section~\ref{secUnitary}.
In contrast to previous works, the \emph{non}linearity 
of our approach gives, in connection with Noether's theorem, 
rise to interesting nontrivial symmetries which deserve further
investigation.

For ease of notation, the presentation focused on systems
with $f=1$ spacial degree of freedom. In fact our
considerations also apply to the general case $f\in\IN$.
Here, phase space $\calM$ consists of all $2f$-tuples
of Cartesian position/momentum observables
$(\IQ_1,\IP_1,\ldots,\IQ_f,\IP_f)$
satisfying commutation relations (\ref{eqVertrel2}).
Correspondingly for the set of abstract scalar fields
(polynomial mappings on $\calM$), we now choose
$\calF = \IC \langle\IQ_1,\IP_1,\ldots,\IQ_f,\IP_f\rangle$,
i.e., the quotient algebra $\IC\langle Q_1,\ldots,P_f\rangle/\Ideal$
with respect to the ideal $\Ideal$ spanned by
$$\big\{ Q_kQ_l-Q_lQ_k \;, \; P_kP_l-P_lP_k  \;, \;
  Q_kP_l-P_lQ_k-i\hbar\delta_{kl} \;:\; 1\leq k,l\leq f \big\} \enspace . $$
Abstract covector fields $\IK\in\calV$ thus become $2f$-tuples
$\IK=(\IK_{q1},\ldots, \IK_{pf})$ s.t.
\vspace*{-1.0ex}
$$
\left.
\begin{aligned}
\IK_{qk}\IQ_l+\IQ_k\IK_{ql} &\quad=\quad \IQ_l\IK_{qk}+\IK_{ql}\IQ_k \\
\IK_{pk}\IP_l+\IP_k\IK_{pl} &\quad=\quad \IP_l\IK_{pk}-\IK_{pl}\IP_k \\
\IK_{qk}\IP_l+\IQ_k\IK_{pl} &\quad=\quad \IP_l\IK_{qk}-\IK_{pl}\IQ_k
\end{aligned} 
\qquad\right\}\qquad \forall 1\leq k,l\leq f
\vspace*{-0.5ex}
$$
where again time-independence
of Planck's constant entered.  Finally for $\IW^*\in\calV^*$,
\vspace*{-1.0ex}
$$
\Theta[\IW^*] \quad := \quad
 \bigg(\!\begin{array}{rr} \Izero_f & +\Ione_f \\[0.7ex] -\Ione_f & \Izero_f
  \end{array}\!\bigg) \cdot
     \left( \begin{array}{c} \IW^*[(1/\Ideal,0/\Ideal,\ldots,0/\Ideal)] \\
    \IW^*[(0/\Ideal,1/\Ideal,\ldots,0/\Ideal)] \\ \vdots \\
    \IW^*[(0/\Ideal,0/\Ideal,\ldots,1/\Ideal)]
       \end{array}  \right) \quad \in \quad \calV
\vspace*{-3.4ex}
$$
is the abstract Noetherian tensor.

\begin{appendix}
\section{Postponed Proof of Theorem~\ref{thNew}}  \label{subsecNew}
First notice that the gradient of a conserved quantity
is an invariant covector field:
$$    \Nabla I[K]=0 \qquad\Longrightarrow\quad   \calL_K(\Nabla I)=0
\enspace . $$
Indeed for any $G\in\calV$, \quad
$$\displaystyle
\calL_K(\Nabla I)[G]
\;\overset{(\ref{eqCovector})}{=}\;
\calL_K(\underbrace{\Nabla I[G]}_{=\calL_G I})
-\Nabla I[\underbrace{\calL_K G}_{\LieBra{K,G}}]
\;=\; \calL_K\calL_G I - \calL_{\LieBra{K,G}} I
\;\overset{(\ref{eqHomomorphism})}{=}\;
\calL_G\underbrace{\calL_K I}_{\Nabla =0}
$$
Thus, according to \textsc{Observation~2.1} in \cite{Coupling},
$$\LieBra{K,G} \;=\; \calL_K(\theta\Nabla I)
  \;\overset{!}{=}\; \theta \calL_K(\Nabla I) \;=\; 0 \enspace . $$

\section{Postponed Proof of Theorem~\ref{thUnitary}}
Let the reader be reminded that an operator $\IU$ on
$\calH$ is called \emph{unitary} iff it is
\\ a) linear and bounded, \quad b) invertible, \quad and
\quad c) satisfies $\IU\IU^*=\II$.

Now the set $\calB(\calH)$ of all bounded linear operators
on $\calH$ is, equipped with operator norm, of course a
Banach algebra and thus in particular a (flat, $C^\infty$) manifold.
Similarly, the set $\calS(\calH)$ of all \emph{symmetric}
bounded linear operators is a (real!) Banach space
and hence a manifold as well.
Let $\calB(\calH)^+$ denote the set
of \emph{invertible} bounded linear operators.
This subset is known to be open in $\calB(\calH)$
and therefore also constitutes a manifold,
cf. e.g. \cite[\textsc{Theorem~10.11}]{Rudin};
similarly, $\calS(\calH)^+$ is open in $\calS(\calH)$
and therefore a manifold, too.

So it holds $\calU(\calH)=f^{-1}(\II)$ for
$f:\calB(\calH)^+\to\calS(\calH)^+$,
$\IA\mapsto\IA\IA^*$. We are going to show that $f$ is
in fact a \emph{submersion} on $\calU(\calH)$.

\begin{definition} \label{defSubmersion}
Let $X$, $Y$ denote Banach manifolds.
Consider a $C^1$ mapping $f:X\to Y$
and $S=f^{-1}(c)\subseteq X$ for some $c\in Y$.
Call $f$ a \defi{submersion} on $S$ if for all $x\in S$,
$\displaystyle f'(x):T_x X\to T_{f(x)}Y$ is surjective
and has a complementable kernel.
\end{definition}

\begin{lemma}
In that case, $S$ is a submanifold of $X$.
\end{lemma}
\begin{iproof}
See \textsc{page 550} in \cite{Manifolds}.
\end{iproof}

Recall that a closed subspace $M$ of a topological vector space
$E$ is called \emph{complementable} if there exists a closed
subspace $N$ of $E$ such that $E=M+N$ and $M\cap N=\{0\}$;
cf. e.g. \cite[\textsc{Definition~4.20}]{Rudin}.

\begin{lemma}
$f:\calB(\calH)^+\to\calS(\calH)^+$, $\IA\mapsto\IA\IA^*$
is continuously differentiable with derivative
$f'(\IA)[\IV]=\IV\IA^*+\IA\IV^*$.

For $\IU\in\calU(\calH)$, the linear map
$f'(\IU):\calB(\calH)\to\calS(\calH)$
is surjective; its kernel
$\calN=\big\{\IB\in\calB(\calH) : \IU\IB^*+\IB\IU^*=0\big\}$
is complemented by
$\calM:=\big\{ \IB\in\calB(\calH) : \IU\IB^*-\IB\IU^*=0\big\}$.
\end{lemma}
\begin{iproof}
Straight-forward calculation yields differentiability of $f$:
$$f(\IA+\IV) - f(\IA) - (\IV\IA^*+\IA\IV^*)
 \quad = \quad \IV\IV^* $$
tends to zero as $\IV\to 0$, even when divided by $\|\IV\|$.

For $\IS\in\calS(\calH)$, let $\IV:=\tfrac{1}{2}\IS\IU\in\calB(\calH)$.
Then, using $\IV^*=\tfrac{1}{2}\IU^*\IS^*$, $\IS^*=\IS$, and $\IU\IU^*=\II$,
it follows that $f'(\IU)[\IV]=\IS$; hence $f'(\IU)$ is surjective.

$\calN\cap\calM=\{0\}$ is trivial. To show $\calN+\calM=\calB(\calH)$,
consider $\IB\in\calB(\calH)$; now verify that
$(\IB-\IU\IB^*\IU)/2\in N$ and $(\IB+\IU\IB^*\IU)/2\in M$.
Since the sum of both yields $\IB$, this concludes the proof.
\end{iproof}

\section{Postponed Proof of Theorem~\ref{thAlgebra}}   \label{subsecAlgebra}
For a) one may presume w.l.o.g. that $q=0$.
Let $d$ denote the degree of $p\not=0$.
According to, e.g., \cite{Ma} there exist symmetric
$(\lfloor\tfrac{d}{2}+1\rfloor\times
\lfloor\tfrac{d}{2}+1\rfloor)$-matrices\footnote{The
famous \textsc{Amitsur-Levitzki-Theorem} states that
this matrix dimension is in fact optimal.}
matrices $A_1,\ldots,A_m$ such that $\hat p(A_1,\ldots,A_m)\not=0$.
By extending the linear mappings $A_j$ from
$\IC^{\lfloor d/2+1\rfloor}$
to $\calH$, the obtained symmetric compact operators
still satisfy $\hat p(\IA_1,\ldots,\IA_m)\not=0$.

\medskip
For b) and c), we are going to algebraically \emph{define} a mapping
$\IC\langle \vec X\rangle\ni p\mapsto p'\in\IC\langle \vec X;\vec V\rangle$
and verify that its image under $\widehat{~}$ coincides with
the derivative of $\hat p$. As the latter is unique on
Hausdorff spaces, this proves the claim.

\begin{definition} \label{defDerivation}
Write $\vec X=(X_1,\ldots,X_m)$ and $\vec V=(V_1,\ldots,V_m)$.
For monomial $p=\prod_{n=1}^d X_{k_n}\in\IC\langle\vec X\rangle$,
its \defi{partial derivative} with respect to $X_l$ is given by
$$   \frac{\partial p}{\partial X_l}[V_l]
  \quad:=\quad \sum_{n:k_n=l}
    \Big( \prod_{s<n}  X_{k_s} \Big) \cdot V_l \cdot
    \Big( \prod_{s>n}  X_{k_s} \Big)
      \quad\in\quad \IC\langle\vec X;V_l\rangle $$
The partial derivative of a linear combination of monomials
is the linear combination of their respective partial derivatives.
The \defi{derivative} and \defi{second derivative} of a polynomial
$p=p(\vec X)\in\IC\langle\vec X\rangle$
are given by
\begin{gather*}
  p' \;= \; p'(\vec X)[\vec V] \;=\; p'[\vec V] \quad:=\quad
     \sum_{l=1}^m   \frac{\partial p}{\partial X_l}[V_l]
      \quad\in\quad \IC\langle\vec X;\vec V\rangle  \\
  p''(\vec X)[\vec V,\vec W] \;:=\;
  \sum_{l=1}^m   \frac{\partial}{\partial X_l}
         \big(p'(\vec X)[\vec V]\big)[W_l]
         \quad\in\quad \IC\langle\vec X;\vec V,\vec W\rangle
\end{gather*}
respectively.
\end{definition}

It's easy to verify the following properties:
\begin{lemma}  \label{lemDerivation}
Let $p,q\in\IC\langle\vec X\rangle$, $\alpha\in\IC$.
\begin{enumerate}
\item  Linearity
  \quad $$\displaystyle (\alpha p+q)'(\vec X)[\vec V]  \;=\;
     \alpha p'(\vec X)[\vec V] + q'(\vec X)[\vec V]$$
\item  Non-commutative product rule
  \quad $$\displaystyle (p\cdot q)'(\vec X)[\vec V] \;=\;
   p(\vec X)[\vec V]\cdot q'(\vec X) + q'(\vec X)[\vec V] \cdot q(\vec X)$$
\item Symmetry of second derivatives
  \quad $\displaystyle p''[\vec V,\vec W]=p''[\vec W,\vec V]$
\item  Chain rule: Let
  $\vec q=(q_1,\ldots,q_m)\in\IC\langle\vec X\rangle^m$. Then
  $$ p\big(\vec q(\vec X)\big)'[\vec V]
   \;=\; p'\big(\vec q(\vec X)\big)\big[\vec q'(\vec X)[\vec V]\big]$$
  with $\vec q'=(q_1',\ldots,q_m')$. In particular,
\begin{equation} \label{eqChainrule}
 \big(p'[\vec q]\big)'[\vec V]
    \;=\; p''[\vec q,\vec V] + p'\big[\vec q'[\vec V]\big] \enspace .
\end{equation}
\end{enumerate}
\end{lemma}

The first two items say that $p\mapsto p'$ is sort of a derivation.
Now finally coming to Claims b) and c), it suffices to
consider monomials; the rest follows from linearity of differentiation.

Let us first remark that the $\widehat{~}$-transform of each
polynomial $p$ is a \emph{continuous} map $\hat p:\calB^m\to\calB$.
Indeed, $p$ is a finite linear combination of products of projections
$(\IA_1,\ldots,\IA_m)\mapsto\IA_j$; the latter are continuous,
and so are products of continuous functions because the
operator norm $\|\cdot\|$ on $\calB$ is submultiplicative.

The proof that $\hat p:\calB^m\to\calB$
is differentiable for every monomial $p\in\IC\langle\vec X\rangle$
proceeds by easy induction on the degree $d$ of $p$,
being obvious for $d\leq 1$. For induction step $d\mapsto d+1$ let
$p\cdot q$ be the product of two monomials $p,q$ of degree
at most $d$ each. By induction hypothesis, both
$\hat p$ and $\hat q$ are differentiable with
respective directional derivatives
$\widehat{p'}(\vec\IA)[\vec\IV]$ and $\widehat{q'}(\vec\IA)[\vec\IV]$,
$\vec\IA,\vec\IV\in\calB^m$.
Recall that this means that
$\hat p(\IA+\IV) - \hat p(\IA) - \widehat{p'}(\IA)[\IV]$
tends to 0 even when divided by $\|\IV\|\to0$
and similarly for $\hat q$. We want to show that
$\widehat{p\cdot q'}+\widehat{p'\cdot q}$
is the derivative of $\widehat{p\cdot q}$.
Because of $\widehat{p\cdot q}=\hat p\cdot\hat q$, it indeed follows
\begin{multline*}
 \widehat{p\cdot q}(\IA+\IV) \;-\; \widehat{p\cdot q}(\IA)
  \;-\; \widehat{p'}(\IA)[\IV]\cdot\hat q(\IA+\IV)
  \;-\; \hat p(\IA)\cdot\widehat{q'}(\IA)[\IV]
\quad = \\  \textstyle
 \underbrace{\big(
 \hat p(\IA+\IV)-\hat p(\IA)-\widehat{p'}(\IA)[\IV]\big)}_{
 \bullet/\|\IV\|\to 0 \text{ as }\IV\to0} \cdot \!\!\!
 \underbrace{\hat q(\IA+\IV)}_{\to\hat q(\IA) \text{ as } \IV\to0}
\!\! +\; \hat p(\IA)\cdot
  \underbrace{\big(\hat q(\IA+\IV)-\hat q(\IA)-\widehat{q'}(\IA)[\IV]\big)}_{
  \bullet/\|\IV\|\to 0 \text{ as }\IV\to0}
\end{multline*}
since $\hat q$ is continuous. As $\|\cdot\|$ satisfies subadditivity
and submultiplicativity, not only the indicated factors
but the whole expression tends to 0 even when divided by $\|\IV\|\to0$.
This shows that $p\cdot q$ is differentiable and its derivative
is the $\widehat{~}$-transform of
$p(\vec X)\cdot q'(\vec X)[\vec V]+p'(\vec X)[\vec V]\cdot q(\vec X)
\in\IC\langle\vec X;\vec V\rangle$ which completes the induction
step and eventually proves Claims b) and c).
\qed

\section{Postponed Proof of Theorem~\ref{thWelldefined}}
\label{subsecWelldefined}
Remember that, for
a (two-sided) ideal $\Ideal$ in some (non-commutative)
algebra $\calA$, the relation
~ $A\relation B \quad:\Leftrightarrow\quad
 A-B\in\Ideal$ ~  satisfies
$$ A\relation B
\qquad\Rightarrow\qquad
A+C\relation B+C \quad\wedge\quad
A\cdot C\relation B\cdot C \quad\wedge\quad
C\cdot A\relation C\cdot B $$
for $A,B,C\in\calA$.
Equivalently: (\ref{eqQuotient}) is well-defined.
A \defi{representative} for $\IA\in\calA/\Ideal$
is some $A\in\calA$ such that $\IA=A/\Ideal$

\medskip
Well-definition of $\calL_{\IK}\IH$ means independence
of the representatives $H\in\IC\langle Q,P\rangle$ for
$\IH\in\IC\langle\IQ,\IP\rangle$ and
similarly $(K_Q,K_P)=\vec K$ for $(\IK_Q,\IK_P)=\IK\in\calV$.
So suppose $\IH=0$ and we have to show that
$H'[(K_Q,K_P)]\relation0$ for each $H\relation 0$.
Indeed linearity allows to presume w.l.o.g.
$H=p\cdot(QP-PQ-i\hbar)\cdot q$
for some $p,q\in\IC\langle Q,P\rangle$. Then
Lemma~\ref{lemDerivation}b) yields
\qquad $\displaystyle H'[\vec K] \;=$
$$ 
      \underbrace{p'[\vec K]\cdot(QP-PQ-i\hbar)\cdot q}_{\relation0}
+ p\cdot\!\!\underbrace{(QP-PQ-i\hbar)'[\vec K]}_{=K_QP+QK_P-K_PQ-PK_Q}\!\!\!
 \cdot q
+ \underbrace{p\cdot(QP-PQ-i\hbar)\cdot q'[\vec K]}_{\relation0}
$$ 
and the middle term is ~$\relation0$ as well because
$(K_Q,K_P)$, being a representative for $(\IK_Q,\IK_P)\in\calV$,
satisfies $\commut{K_Q,P}+\commut{Q,K_P}\relation0$ according
to Definition~\ref{defVector}.

Derivatives in direction of vector fields thus
basically 'commute' with taking factors w.r.t. $\Ideal$:
$$  (H/\Ideal)'[\vec K/\Ideal] \quad = \quad
  \big(H'[\vec K]\big)/\Ideal
    \qquad \text{ for } (\vec K)/\Ideal\in\calV $$
For \emph{partial} derivatives, this does in general not hold:
Take $H_1:=QP-PQ$, $H_2=i\hbar$, and $V:=Q$;
then $H_1/\Ideal=H_2/\Ideal$ but
$$ \Big(\frac{\partial H_1}{\partial Q}[V]\Big)/\Ideal
  \;=\; (QP-PQ)/\Ideal \;=\; (i\hbar)/\Ideal  \quad\not=\quad
  0/\Ideal \;=\; \Big(\frac{\partial H_2}{\partial Q}[V]\Big)/\Ideal
\enspace . $$
If however $VP\equiv PV$, then $(V,0)/\Ideal$ belongs to $\calV$ and
$\Big(\frac{\partial H}{\partial Q}[V]\big)/\Ideal=\big(H'[(V,0)]\big)/\Ideal$
is independent of $H$ as some representative for $H/\Ideal$; same for
$\Big(\frac{\partial H}{\partial P}[V]\big)/\Ideal=\big(H'[(0,V)]\big)/\Ideal$
whenever $VQ\equiv QV$. In particular for $V=1$, we therefore have

\begin{lemma} \label{lemHilfssatz}
Let $\IH\in\calF$ with $\IH=H/\Ideal$. Then
$$ \frac{\partial\IH}{\partial\IQ}
\quad:=\quad \Big(\frac{\partial H}{\partial Q}[1]\Big)/\Ideal, \qquad
\frac{\partial\IH}{\partial\IP}
\quad:=\quad \Big(\frac{\partial H}{\partial P}[1]\Big)/\Ideal
$$
is well-defined. Furthermore it holds
\begin{equation} \label{eqHilfssatz}
\frac{i}{\hbar}\commut{\IH,\IQ} \quad=\quad +\frac{\partial\IH}{\partial\IP},
\qquad
\frac{i}{\hbar}\commut{\IH,\IP} \quad=\quad -\frac{\partial\IH}{\partial\IQ}
\end{equation}
\end{lemma}

Condition (\ref{eqVector}) for $(\IK_Q,\IK_P)\in\calV$
may thus be rewritten as
$$  \frac{\partial\IK_Q}{\partial\IQ} \quad=\quad -
    \frac{\partial\IK_P}{\partial\IP} $$
which resembles the \textsf{Cauchy-Riemann} equation for the
complex function $f(q+ip)=k_q(q,p)-ik_p(q,p)$
to be differentiable.

\begin{proof}{Lemma~\ref{lemHilfssatz}}
Let $\IH=H/\Ideal\in\calF$. For linearity reasons, it suffices to prove
$\tfrac{i}{\hbar}\commut{\IH,\IQ}=\frac{\partial\IH}{\partial\IP}$
for monomials $H$. We proceed by induction on the degree of $H$,
cases $H=1$, $H=Q$, and $H=P$ being obvious.
So let $H=H_1\cdot H_2$ with monomials $H_1,H_2$ of lower degree,
$\IH_1=H_1/\Ideal$, $\IH_2=H_2/\Ideal$. Then
\begin{eqnarray*}
\frac{\partial\IH}{\partial\IP}
&=& \Big(\frac{\partial (H_1\cdot H_2)}{\partial P}[1]\Big)/\Ideal \\
&\overset{\ref{lemDerivation}b}{=}&
  \Big(\frac{\partial H_1}{\partial P}[1]\cdot H_2
      +H_1\cdot\frac{\partial H_2}{\partial P}[1]\Big)/\Ideal
\quad\overset{(\ref{eqQuotient})}{=}\quad
  \frac{\partial\IH_1}{\IP}\cdot\IH_2 + \IH_1\cdot\frac{\partial\IH_2}{\IP} \\
&\overset{(*)}{=}&
  \tfrac{i}{\hbar}\commut{\IH_1,\IQ}\cdot\IH_2 + \IH_1\cdot
  \tfrac{i}{\hbar}\commut{\IH_2,\IQ}
\quad=\quad \tfrac{i}{\hbar}\commut{\IH_1\cdot\IH_2,\IQ}
\end{eqnarray*}
where at $(*)$ the inductive presumption entered.
\end{proof}

\bigskip
Next claim is that $\LieBra{\IK,\IG}$ belongs to $\calV$
for $\IK,\IG\in\calV$. To this end, take corresponding
representatives $(K_Q,K_P)$ and $(G_Q,G_P)$ --- which ones
doesn't matter as we have just shown --- and verify
that the representative $\big(G_Q'[K]-K_Q'[G],G_P'[K]-K_P'[G]\big)$
for $\LieBra{\IK,\IG}$ satisfies
\begin{eqnarray*}
\lefteqn{\commut{G_Q'[K]-K_Q'[G],P}\;+\;\commut{Q,G_P'[K]-K_P'[G]}} \\
&\overset{\ref{defDerivation}}{=}&
    \commut{\frac{\partial G_Q}{\partial Q}[K_Q],P}
  + \commut{\frac{\partial G_Q}{\partial P}[K_P],P}
  - \commut{\frac{\partial K_Q}{\partial Q}[G_Q],P}
  - \commut{\frac{\partial K_Q}{\partial P}[G_P],P}  \\
&+& \commut{Q,\frac{\partial G_P}{\partial Q}[K_Q]}
  + \commut{Q,\frac{\partial G_P}{\partial P}[K_P]}
  - \commut{Q,\frac{\partial K_P}{\partial Q}[G_Q]}
  - \commut{Q,\frac{\partial K_P}{\partial P}[G_P]} \\
&\overset{\ref{lemDerivation}b}{=}&
    \frac{\partial}{\partial Q}\big(\commut{G_Q,P}\big)[K_Q]
  \quad+\quad \frac{\partial}{\partial P}\big(\commut{G_Q,P}\big)[K_P]
         -\commut{G_Q,K_P}\\
&-& \frac{\partial}{\partial Q}\big(\commut{K_Q,P}\big)[G_Q]
 \quad-\quad \frac{\partial}{\partial P}\big(\commut{K_Q,P}\big)[G_P]
         +\commut{K_Q,G_P}\\
&+& \frac{\partial}{\partial Q}\big(\commut{Q,G_P}\big)[K_Q]-\commut{K_Q,G_P}
 \quad+\quad \frac{\partial}{\partial P}\big(\commut{Q,G_P}\big)[K_P]  \\
&-& \frac{\partial}{\partial Q}\big(\commut{Q,K_P}\big)[G_Q]+\commut{G_Q,K_P}
 \quad-\quad \frac{\partial}{\partial P}\big(\commut{Q,K_P}\big)[G_P]
\\  
&\overset{\ref{lemDerivation}a}{=}&
    \frac{\partial}{\partial Q}\big(\commut{G_Q,P}+\commut{Q,G_P}\big)[K_Q]
  + \frac{\partial}{\partial P}\big(\commut{G_Q,P}+\commut{Q,G_P}\big)[K_P] \\
&-& \frac{\partial}{\partial Q}\big(\commut{K_Q,P}+\commut{Q,K_P}\big)[G_Q]
  - \frac{\partial}{\partial P}\big(\commut{K_Q,P}+\commut{Q,K_P}\big)[G_P] \\
&\overset{\ref{defDerivation}}{=}&
  \big(\underbrace{\commut{G_Q,P}+\commut{Q,G_P}}_{\relation0}\big)'[K]
 - \big(\underbrace{\commut{K_Q,P}+\commut{Q,K_P}}_{\relation0}\big)'[G]
\quad \relation \quad 0 \enspace .
\end{eqnarray*}
Indeed, Lemma~\ref{lemDerivation}b) implies
$$ \frac{\partial}{\partial Q}\commut{A,B}[V]
\quad=\quad \commut{\frac{\partial A}{\partial Q}[V],B}
  + \commut{A,\frac{\partial B}{\partial Q}[V]} $$
and for $B=Q$, the last term is equal to $\commut{A,V}$
whereas it vanishes for $B=P$.

\bigskip
The mapping $\calV\ni\IK\mapsto\calL_{\IK}$ is a Lie algebra homomorphism
because, for $\IK=K/\Ideal, \IG=G/\Ideal\in\calV$, and $H=H/\Ideal\in\calF$,
\begin{eqnarray*}
\calL_{\IK}\calL_{\IG}-\calL_{\IG}\calL_{\IK}:\IH & \mapsto &
  \Big( \big(H'[G]\big)'[K]-\big(H'[K]\big)'[G] \Big)/\Ideal  \\
& \underset{(\ref{eqChainrule})}{\overset{\ref{lemDerivation}d)}{=}} &
  \Big( H''\big[G,K\big] + H'\big[G'[K]\big] - H''\big[K,G\big]
      - H'\big[K'[G]\big] \Big)/\Ideal \\
& \underset{a+c}{\overset{\ref{lemDerivation}}{=}} &
  \Big( H'\big[G'[K]-K'[G]\big] \Big)/\Ideal
  \quad=\quad \calL_{\LieBra{\IK,\IG}} \IH \enspace .
\end{eqnarray*}
This is furthermore injective as can be seen by evaluating
$\calL_{\IK}\IH=\cal_{\IG}\IH$ on $\IH:=\IQ$ and on $\IH:=\IP$.
In particular, $\LieBra{\,\cdot\,,\,\cdot\,}$ satisfies
antisymmetry and Jacobi's identity.

\section{Postponed Proof of Theorem~\ref{thImplectic}}  \label{subsecImplectic}
Let us first emphasize the importance of ideal $\Ideal$
by omitting it, that is, by considering
$$ \tilde\Theta[W^*] \quad := \quad
   \bigg(\!\begin{array}{rr} 0 & +1 \\[0.7ex] -1 & 0
  \end{array}\!\bigg) \cdot
     \bigg( \begin{array}{c} W^*[(1,0)] \\[0.7ex]
       W^*[(0,1)] \end{array}
     \bigg)   \qquad\text{ on }\qquad
\tilde\calF=\IC\langle Q,P\rangle \enspace . $$
This linear mapping from $\tilde\calV^*\to\tilde\calV$ is,
in spite of its suggestive writing, not even antisymmetric:
For $H:=Q^2\in\tilde\calF$ and $F:=PQP\in\tilde\calF$,
\begin{eqnarray*}
\lefteqn{
\Nabla F\big[\tilde\Theta[\Nabla H]\big] +
\Nabla H\big[\tilde\Theta[\Nabla F]\big]  } \\
&=&
\Nabla F\big[(0,-2Q)\big] + \Nabla H\big[(QP+PQ,-P^2)\big] \\
&=& \big(-2Q^2P-2PQ^2\big) + \big(Q(QP+PQ)+(QP+PQ)Q\big)  \\
&=& 2QPQ-Q^2P-PQ^2 \quad \not= \quad 0 \enspace .
\end{eqnarray*}

\bigskip
Now returning to the proof of Theorem~\ref{thImplectic},
$\IW^*\big[(1/\Ideal,0/\Ideal)\big]$ is well-defined
because $\IK=(1/\Ideal,0/\Ideal)$ satisfies
$\commut{\IK_Q,\IP}=\commut{\IK_P,\IQ}$,
thus belongs to $\calV$ on which
$\IW^*:\calV\to\calF$ operates.

Next we exploit that according to Definition~\ref{defNew},
$\calV^*$ consists of abstract gradients
(i.e., \emph{closed} covector fields) only.
Namely to show $\Theta[\Nabla\IH]\in\calV$, take
$\IH=H/\Ideal$ and compute for
$K=(K_Q,K_P):=\tilde\Theta[\Nabla H]$
\begin{eqnarray*}
\commut{K_Q,P}+\commut{Q,K_P}
&=& \commut{\frac{\partial H}{\partial P}[1],P}
  -\commut{Q,\frac{\partial H}{\partial Q}[1]} \\
&\overset{\ref{lemHilfssatz}}{\relation}&
  \commut{ \tfrac{i}{\hbar}\commut{H,Q} , P}
  +\commut{Q, \tfrac{i}{\hbar}\commut{H,P}} \\
&\overset{(*)}{=}& - \commut{ \tfrac{i}{\hbar}\commut{Q,P} , H }
\quad\relation\quad \commut{ 1,H } \quad=\quad 0
\end{eqnarray*}
where at $(*)$, Jacobi's identity was used:
$$ \commut{ \commut{A,B},C } \;+\; \commut{ \commut{B,C} ,A}
  \;+\; \commut{\commut{B,A},C} \quad=\quad 0 \enspace . $$

$\Theta$ obviously satisfies linearity;
for proving that it is furthermore antisymmetric and Noetherian,
the following tool turns out to be quite useful:

\begin{proposition}  \label{proHilfssatz}
For $m,n,M,N\in\IN_0$, $\alpha,\beta\in\IC$, ~
$A:=\alpha P^mQ^n$ and $B:=\beta P^MQ^N$ satisfy
$$  \frac{\partial A}{\partial Q}\Big[\frac{\partial B}{\partial P}[1]
\Big] \quad\relation\quad \frac{\partial B}{\partial P}
\Big[\frac{\partial A}{\partial Q}[1]\Big] \enspace . $$
\end{proposition}

Notice that both terms in general coincide only with respect to the
relation $\relation$ induced by identifying $QP-PQ$ with $i\hbar$;
consider, e.g., $A=Q^2$ and $B=P^3$.
Furthermore, the particular form of $A$ and $B$ (with all
$P$'s to the left and $Q$s to the right) is important;
consider, e.g., $A=Q^2$ and $B=PQP$. The latter results
from the fact that, say, $\frac{\partial B}{\partial P}[1]$
is usually not the (representative of the) first component
of a vector field and thus
partial derivative $\frac{\partial A}{\partial Q}$
in this direction not necessarily independent of
changing representatives for $A$.

Nevertheless, Proposition~\ref{proHilfssatz} helps
calculating, for $\IF,\IH\in\calF$,
$\Nabla\IF\big[\Theta[\Nabla\IH]\big]+\Nabla\IH\big[\Theta[\Nabla\IF]\big]
=0$. Indeed, one may presume w.l.o.g. that
$\IF=F/\Ideal$ and $\IH=H/\Ideal$ for $F=P^mQ^n$ and $H=P^MQ^N$
because of (bi-)linearity and since every monomial in $\calH$
can be brought to this form; cf. Lemma~\ref{lemVertrel}.
Then,
\begin{eqnarray*}
\lefteqn{
\Nabla\IF\big[\Theta[\Nabla\IH]\big] \;+\;
\Nabla\IH\big[\Theta[\Nabla\IF]\big] } \\[1ex]
&\overset{\ref{defSymplectic}}{=}&
\frac{\partial F}{\partial Q}\Big[\frac{\partial H}{\partial P}[1]\Big]/\Ideal
\;-\;
\frac{\partial F}{\partial P}\Big[\frac{\partial H}{\partial Q}[1]\Big]/\Ideal
\;+\;
\frac{\partial H}{\partial Q}\Big[\frac{\partial F}{\partial Q}[1]\Big]/\Ideal
\;-\;
\frac{\partial H}{\partial P}\Big[\frac{\partial F}{\partial P}[1]\Big]/\Ideal
\\ &\overset{\ref{proHilfssatz}}{=}& 0 \;+\; 0 \enspace .
\end{eqnarray*}

We now prove that the generalization of usual Poisson brackets
$$ (\IF,\IH) \;\mapsto\; \Nabla\IH[\Theta\Nabla\IF] $$
turns the abstract scalar fields into a Lie algebra.
According to \cite[\textsc{Theorem~4.5}]{Benno}, this is
equivalent (among others) to $\Theta$ being Noetherian
and furthermore to
$$ \Theta\Nabla\big[\Nabla\IH[\Theta\Nabla\IF]\big]
 \quad=\quad \LieBra{\Theta\Nabla\IF,\Theta\Nabla\IH} $$
for all closed covectors $\Nabla\IF,\Nabla\IH\in\calV^*$;
cf. \textsc{Equation~(2.10)} in \cite{Coupling}.
So let, again without loss of generality,
$\IF=F/\Ideal$ and $\IH=H/\Ideal$ with $F=P^mQ^n$ and $H=P^MQ^N$.
Consider
$A:=\frac{\partial H}{\partial Q}\Big[\frac{\partial F}{\partial P}[1]\Big]
\,-\,\frac{\partial H}{\partial P}\Big[\frac{\partial F}{\partial Q}[1]\Big]$,
that is,
$A/\Ideal = \Nabla\IH[\Theta\Nabla\IF]$; then
the first component of $(\IK_Q,\IK_P)=\Theta\Nabla\big[
\Nabla\IH[\Theta\Nabla\IF]\big]$
equals
\begin{eqnarray*}
\IK_Q &\overset{\ref{defSymplectic}}{=}&
\frac{\partial}{\partial P} A[1]/\Ideal \\
&\overset{\ref{lemDerivation}d}{\underset{(\ref{eqChainrule})}{=}}&
\frac{\partial^2 H}{\partial P\partial Q}\Big[1,\frac{\partial F}{\partial P}
  [1]\Big]/\Ideal
\;+\; \frac{\partial H}{\partial P}\Big[\frac{\partial}{\partial Q}
  \Big(\frac{\partial F}{\partial P}[1]\Big)[1]\Big]/\Ideal  \\
&-& \frac{\partial^2 H}{\partial P\partial P}\Big[ 1,
  \frac{\partial F}{\partial Q}[1]\Big]/\Ideal
\;-\; \frac{\partial H}{\partial P}\Big[\frac{\partial}{\partial P}\Big(
  \frac{\partial F}{\partial Q}[1]\Big)[1]\Big]/\Ideal   \\
&\overset{(*)}{=}&
  \frac{\partial}{\partial Q}\Big(\frac{\partial H}{\partial P}[1]\Big)
  \Big[\frac{\partial F}{\partial P}[1]\Big]/\Ideal
\;+\; \frac{\partial H}{\partial Q}\Big[\frac{\partial}{\partial P}
  \underbrace{\Big(\frac{\partial F}{\partial P}[1]\Big)}_{=mP^{m-1}Q^n=:B}
  [1]\Big]/\Ideal \\
&-& \frac{\partial}{\partial P}\Big(\frac{\partial H}{\partial P}[1]\Big)
  \Big[\frac{\partial F}{\partial Q}[1]\Big]/\Ideal
\;-\;\frac{\partial H}{\partial P}\Big[\frac{\partial}{\partial Q}
  \underbrace{\Big(\frac{\partial F}{\partial P}[1]\Big)}_{=B}
    [1]\Big]/\Ideal
\\  
&\overset{\ref{proHilfssatz}}{=}&
  \frac{\partial}{\partial Q}\Big(\frac{\partial H}{\partial P}[1]\Big)
  \Big[\frac{\partial F}{\partial P}[1]\Big]/\Ideal
\;+\; \frac{\partial}{\partial P}\Big(\frac{\partial F}{\partial P}
  [1]\Big)\Big[\frac{\partial H}{\partial Q}[1]\Big]/\Ideal \\
&-& \frac{\partial}{\partial P}\Big(\frac{\partial H}{\partial P}[1]\Big)
  \Big[\frac{\partial F}{\partial Q}[1]\Big]/\Ideal
\;-\; \frac{\partial}{\partial Q}\Big(\frac{\partial F}{\partial P}[1]\Big)
  \Big[\frac{\partial H}{\partial P}[1]\Big]/\Ideal  \\
&\overset{\ref{defSymplectic}}{=}&
  \Big(\frac{\partial\IH}{\partial P}[1]\Big)'\big[\Theta\Nabla\IF\big]
  \;-\; \Big(\frac{\partial\IF}{\partial P}[1]\Big)'\big[\Theta\Nabla\IH\big]
\end{eqnarray*}
which is the first component of
$\LieBra{\Theta[\Nabla\IF],\Theta[\Nabla\IH]}$;
that second components agree as well can be verified quite similarly.
Let us emphasize that at $(*)$, we used
$$ \frac{\partial}{\partial X_k}\Big(\frac{\partial H}{\partial X_l}
  [1]\Big)\big[V\big]
\overset{\ref{lemDerivation}d}{\underset{(\ref{eqChainrule})}{=}}
  \frac{\partial^2 H}{\partial X_k\partial X_l}\big[V,1\big]
  \;+\; \Big(\frac{\partial H}{\partial X_l}[1]\Big)'
   \Big[\underbrace{\frac{\partial 1}{\partial X_k}[V]}_{=0}\Big] $$
for $H,V\in\IC\langle X_1,\ldots,X_m\rangle$.

That $\Theta[\Nabla\IH]$ agrees with $\IK$ according to
(\ref{eqHeisenberg}) follows from Lemma~\ref{lemHilfssatz}.

\section{Postponed Proof of Proposition~\ref{proHilfssatz}}
First thing to notice is that because of linearity,
one may presume $\alpha=\beta=1$.

Next, the claim follows from $A=Q^n$ and $B=P^M$ via induction. Indeed,
once it holds for $A$ and $B$, we have for $\tilde B=B\cdot Q$:
$$ \frac{\partial (B\cdot Q)}{\partial P}\Big[\frac{\partial A}{\partial Q}
  [1]\Big]
  \overset{\ref{lemDerivation}b}{=}
  \frac{\partial B}{\partial P}\Big[\frac{\partial A}{\partial Q}[1]\Big]
  \cdot Q \overset{\text{I.H.}}{\relation}
  \frac{\partial A}{\partial Q}\Big[\frac{\partial B}{\partial P}[1]\Big]
  \cdot Q \overset{(*)}{=} \frac{\partial A}{\partial Q}\Big[
  \frac{\partial B}{\partial P}[1]\cdot Q\Big] $$
where at $(*)$ we used
$$ \frac{\partial A}{\partial Q}[H\cdot Q]
 \;\overset{\ref{defDerivation}}{=}\;
   \sum_{k=1}^n  P^m Q^{k-1} (H\cdot \underbrace{Q) Q^{n-k}}_{=Q^{n-k}\cdot Q}
 \;=\; \sum_{k=1}^n  P^m Q^{k-1} H Q^{n-k} \cdot Q
 \;=\; \frac{\partial A}{\partial Q}[H]\cdot Q \enspace ; $$
the induction step proceeds similarly for $\tilde A=P\cdot A$.

It thus remains to prove
\begin{equation} \label{eqProposition}
 \frac{\partial(Q^n)}{\partial Q}\big[m\cdot P^{m-1}\big]
  \quad\relation\quad \frac{\partial(P^m)}{\partial P}\big[n\cdot Q^{n-1}\big]
\end{equation}
for any $n,m\in\IN$. To this end, we need
the commutation properties of $\IQ^n$ and $\IP^m$.
For $n=1=m$, they are revealed by (\ref{eqVertrel});
and based on that, induction yields:

\begin{lemma} \label{lemVertrel}
Consider $\IQ^n,\IP^m\in\calF$. Then
\begin{gather*}
  \IQ^n\IP-\IP\IQ^n \;=\; n\hbar i \IQ^{n-1}, \qquad
  \IQ\IP^m-\IP^m\IQ \;=\; m\hbar i \IP^{m-1} , \\
    \IQ^n\IP^m-\IP^m\IQ^n \;=\; \sum_{r=1}^{\min(n,m)}
    \binom{m}{r} \binom{n}{r} r! (i\hbar)^r \IP^{m-r} \IQ^{n-r}
\end{gather*}
\end{lemma}

With the agreement that $\binom{k}{r}=0$ for $k<r$,
we may omit the minimum and let the sum range up to $n$ or to $m$
whatever seems preferable.
Now turning to the proof of (\ref{eqProposition}):

\begin{eqnarray*}
\lefteqn{ \frac{\partial(Q^n)}{\partial Q}\big[m\cdot P^{m-1}\big]
 \;-\; \frac{\partial(P^m)}{\partial P}\big[n\cdot Q^{n-1}\big] } \\
&=& m\sum_{k=1}^n \underbrace{Q^{k-1} P^{m-1}}_{\textstyle
  \overset{\ref{lemVertrel}}{\relation}
   P^{m-1}Q^{k-1} + \sum_{r=1}^{m-1} \binom{m-1}{r}\binom{k-1}{r}
    r! (i\hbar)^r P^{m-1-r} Q^{k-1-r}\hspace*{-53ex}}            Q^{n-k}  \\
&-& n\sum_{l=1}^m P^{l-1} \underbrace{Q^{n-1}P^{m-l}}_{\textstyle
   \overset{\ref{lemVertrel}}{\relation}
    P^{m-l}Q^{n-1}+\sum_{s=1}^{n-1} \binom{n-1}{s}\binom{m-l}{s}
     s! (i\hbar)^s P^{m-l-s} Q^{n-1-s}\hspace*{-48ex}}
   \hspace*{43ex}
\end{eqnarray*}
\begin{eqnarray*}
&\relation&
  \underbrace{m\sum_{k=1}^n P^{m-1}Q^{n-1}-n\sum_{l=1}^m P^{m-1} Q^{n-1}}_{=0}
\\ &+& m  \sum_{r=1}^{m-1}  \sum_{k=1}^n
  \binom{m-1}{r} \binom{k-1}{r} r! (i\hbar)^r P^{m-1-r} Q^{n-1-r}
\\ &-& n  \sum_{s=1}^{n-1}  \sum_{l=1}^m
  \binom{n-1}{s} \binom{m-l}{s} s! (i\hbar)^s P^{m-1-s} Q^{n-1-s}
\enspace .
\end{eqnarray*}
Here, both sum ranges for $r$ and $s$ may be cut off at $\min(m-1,n-1)$
since for higher indices, the corresponding binomial coefficients
are zero anyway. Collecting term thus yields
$$ 
\sum_{t=1}^{\scriptscriptstyle\min(n,m)-1\hspace*{-2.2ex}}
     \big( m\cdot\binom{m-1}{t}\sum_{k=1}^n \binom{k-1}{t}
\;-\; n\cdot\binom{n-1}{s}\sum_{l=1}^m \binom{m-l}{t} \big)
              \; t! (i\hbar)^t P^{m-1-t} Q^{n-1-t}
$$ 
which vanishes because the well-known properties of binomial coefficients
$$ \sum_{j=0}^J\binom{j}{N} \quad\overset{(*)}{=}\quad \binom{J+1}{N+1}
 \qquad\text{ and }\qquad
   \binom{J}{N} \quad\underset{(*)}{=}\quad \frac{J}{N}
  \cdot\binom{J-1}{N-1} $$
yield
\begin{eqnarray*}
m\cdot\binom{m-1}{t}\sum_{k=1}^n \binom{k-1}{t}
&\overset{(*)}{=}& m\cdot\binom{m-1}{t}\cdot \binom{n}{t+1} \\
&\underset{(*)}{=}& m\cdot\binom{m-1}{t}\cdot\frac{n}{t+1}\cdot\binom{n-1}{t}
\\[3ex]
n\cdot\binom{n-1}{t}\sum_{l=1}^m \binom{m-l}{t}
&\overset{(*)}{=}& n\cdot\binom{n-1}{t}\cdot\binom{m}{t+1} \\
&\underset{(*)}{=}& n\cdot\binom{n-1}{t}\cdot\frac{m}{t+1}\cdot\binom{m-1}{t}
\enspace . \qed
\end{eqnarray*}
\end{appendix}

\vspace*{-0.5ex}
{ \small
\begin{thebibliography}{21}
\vspace*{-0.5ex}
\itemsep0.1pt
\bibitem{Mechanics}
  \textsc{Abraham, R., and J.E. Marsden}:
  "Foundations of Mechanics" ~2nd Edition,
    Benjamin Cummings Publishing Company (1978).
\bibitem{Symmetries}
  \textsc{Chaichian, M., and R. Hagedorn}:
  "Symmetries in Quantum Mechanics",
    Institute of Physics Publishing (1998).
\bibitem{Chernoff}
  \textsc{Chernoff, P.R., and J.E. Marsden}:
  "Properties of Infinite Dimensional Hamiltonian Systems",
  Springer Lecture Notes in Mathematics {\bf 425} (1974).
\bibitem{Manifolds}
  \textsc{Choquet-Bruhat, Y., and C. deWitt-Morette, and M. Dillard-Bleick}:
    "Analysis, Manifolds, and Physics", North-Holland (1991).
\bibitem{Cirelli}
  \textsc{Cirelli, R., and L. Pizzocchero}:
  "On the integrability of quantum mechanics as an infinite dimensional
  Hamiltonian system", pp.1057-1080 in \emph{Nonlinearity}
  {\bf 3} (1990).
\bibitem{DPG}
  \textsc{Faddeev, L.D.}:
  "How we understand `quantization' a hundred years
  after Max Planck", pp.689-690 in \emph{Physikalische Bl\"{a}tter}
    {\bf 52} (1996).
\bibitem{Benno}
  \textsc{Fuchssteiner, B.}:
  "Hamiltonian Structure and Integrability",
  pp.211-256 in \emph{Nonlinear Equations in the Applied Sciences},
    Academic Press (1992).
\bibitem{Spin}
  \textsc{Fuchssteiner, B.}:
  "An alternative dynamical description of Quantum Systems",
  in \emph{Quantum Groups and Related Topics}, Kluwer (1992).
\bibitem{Coupling}
  \textsc{Fuchssteiner, B.}:
  "Coupling of Completely integrable Systems:
  The Perturbation Bundle", pp.125-138 in \emph{Applications of Analytic
  and Geometric Methods to Nonlinear Differential Equations},
  Kluwer (1993).
\bibitem{Computer}
  \textsc{Fuchssteiner, B., and S. Ivanov, and W. Wiwianka}:
  "Algorithmic determination of infinite dimensional symmetry groups
  for integrable systems", pp.91-100 in
  \emph{Mathematical and Computer Modelling} {\bf 25(8)} (1997).
\bibitem{Gardner}
  \textsc{Gardner, C.S.}:
  "The Korteweg-de Vries Equation as a Hamiltonian System",
  pp.1548-1551 in \emph{J. Math. Phys.} {\bf 12} (1971).
\bibitem{Heslot}
  \textsc{Heslot, A.}: "Quantum mechanics as a classical theory",
  pp.1341-1348 in \emph{Phys. Rev. D} {\bf 31(6)} (1985).
\bibitem{Kupershmidt}
  \textsc{Kupershmidt, B.A.}: "Quantum mechanics as an integrable system",
  pp.136-138 in \emph{Phys. Lett. A} {\bf 109} (1985).
\bibitem{Landsman}
  \textsc{Landsman, N.P.}: "Mathematical Topics Between Classical and
  Quantum Mechanics", Springer (1991).
\bibitem{Ma}
  \textsc{Ma, W., and M.L. Racine}:
    "Minimal identities of symmetric matrices",
  pp.171-192 in \emph{Amer. Math. Soc.} {\bf 320} (1990).
\bibitem{Marsden}
  \textsc{Marsden, J.E., and A. Weinstein}:
  "Reduction of Symplectic Manifolds with Symmetry",
  \emph{Reports on Math. Phys.} {\bf 5(1)} (1974).
\bibitem{vNeumann}
  \textsc{von Neumann, J.}:
  "Mathematical Foundations of Quantum Mechanics",
  Springer (1932), translation Princeton University Press (1955).
\bibitem{Putnam}
  \textsc{Putnam, C}: "Commutation Properties of Hilbert Space Operators",
  Springer (1967).
\bibitem{Reed}
  \textsc{Reed, M., and B. Simon}: "Functional Analysis",
  \emph{Methods of Modern Mathematical Physics} {\bf I},
  Academic Press (1972).
\bibitem{Rudin}
  \textsc{Rudin, W.}: "Functional Analysis", McGraw-Hill (1991).
\bibitem{KdV}
  \textsc{Zakharov, V.E., and L.D. Faddeev}:
  "Korteweg-de Vries equation: a completely integrable system",
  pp.280-287 in \emph{Functional Analysis Appl.} {\bf 5} (1971).
\end{thebibliography}
}
\end{document}